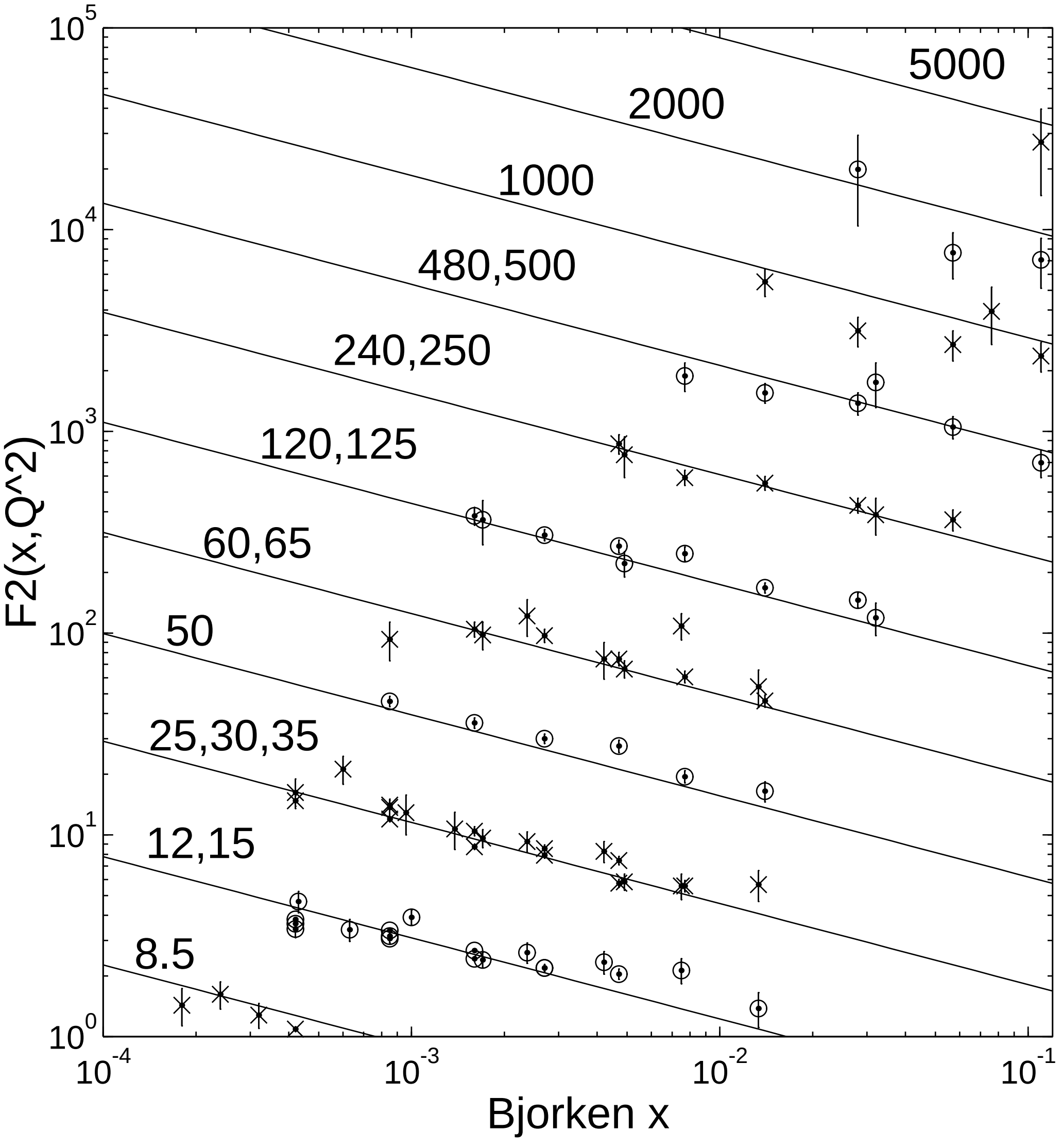

Fig. 1

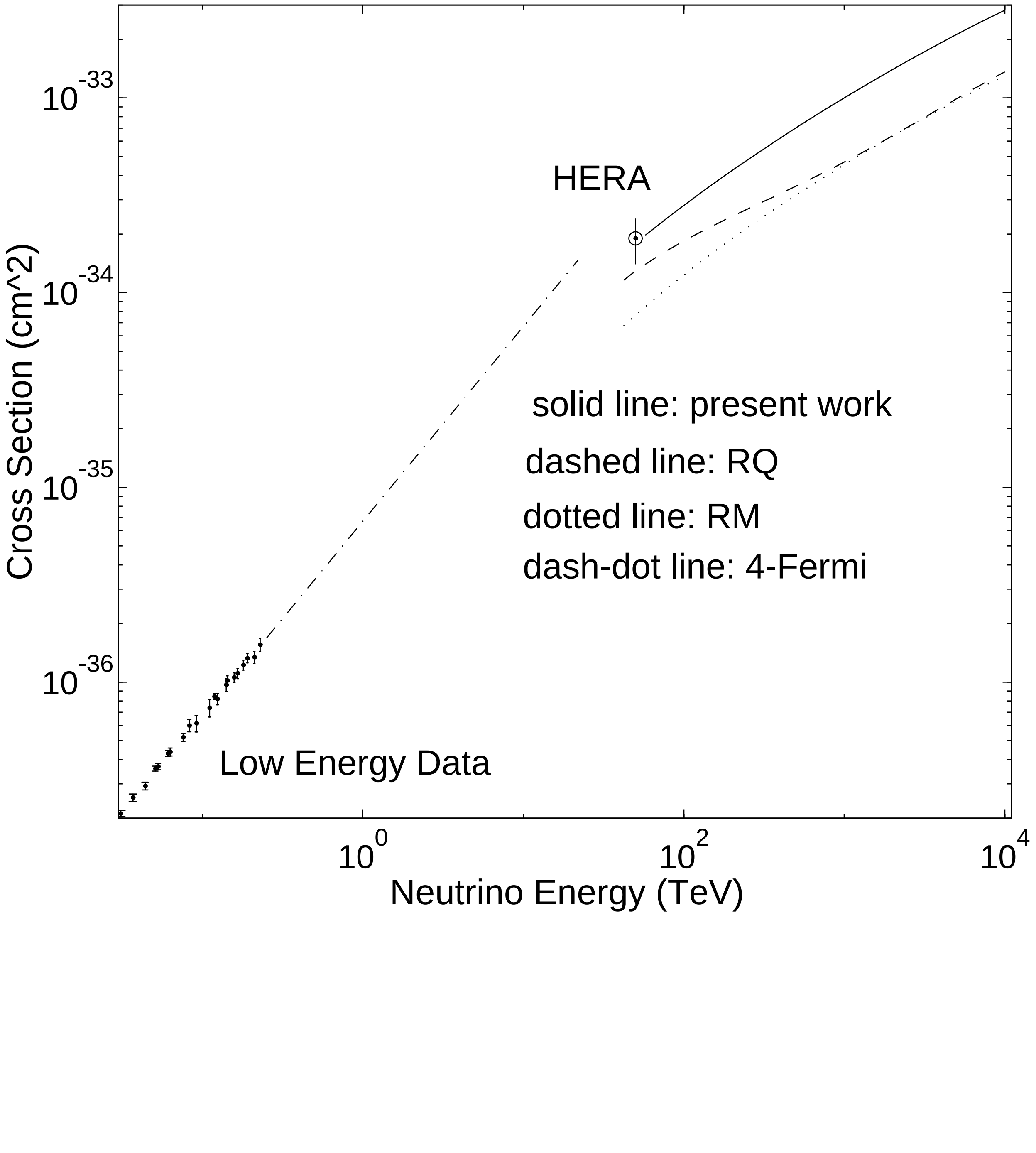

Fig. 2

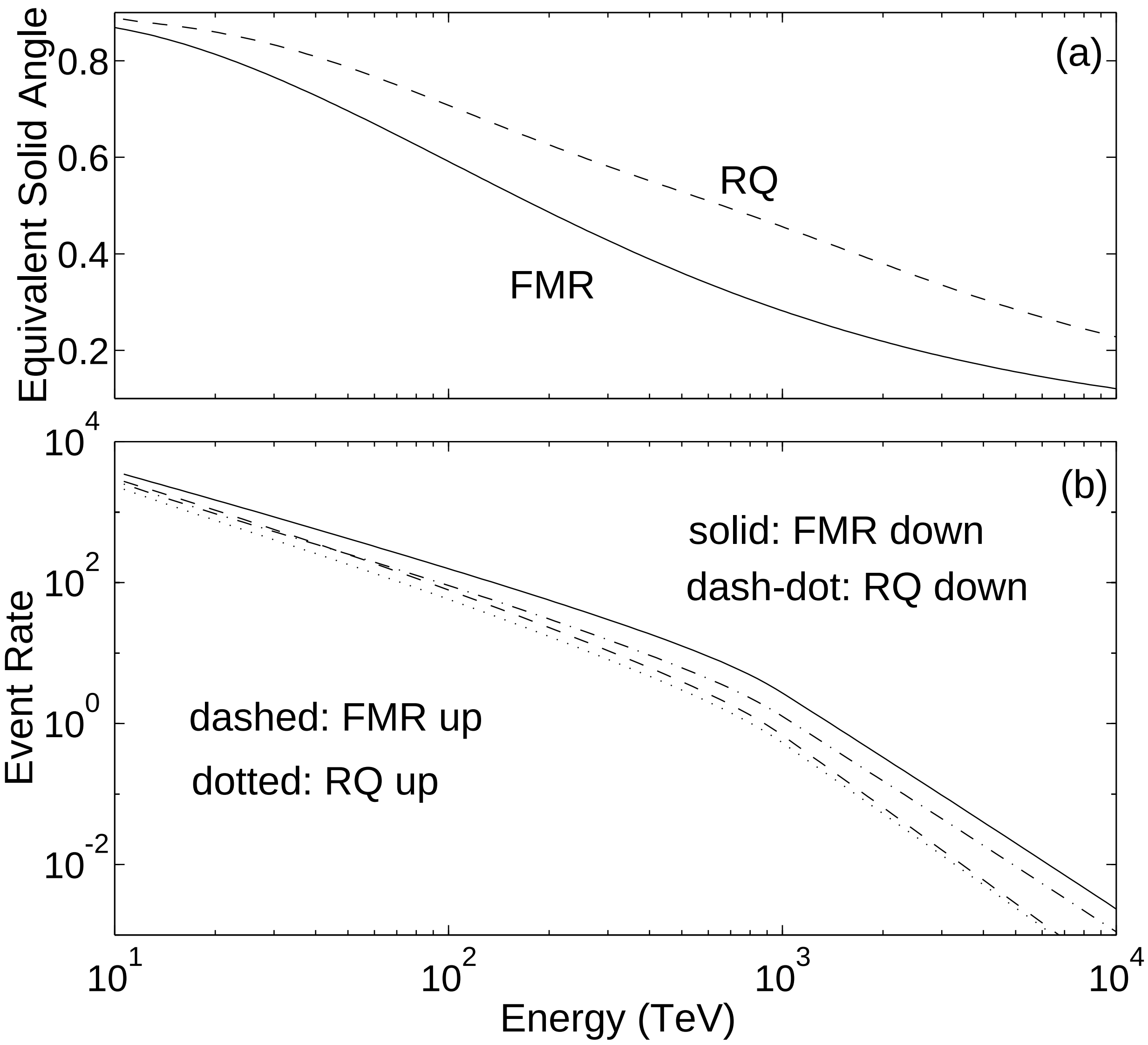

Fig. 3

# Small-X Quarks from HERA Predict the Ultra High Energy Neutrino-Nucleon Cross Section

George M. Frichter, Douglas W. McKay and John P. Ralston

*Department of Physics and Astronomy, and*

*Kansas Institute for Theoretical and Computational Science,*

*University of Kansas,*

*Lawrence, KS 66045, USA*

## Abstract

New structure function data at small Bjorken $x$ from HERA are used along with next-to-leading order QCD evolution to predict a cross section for charged-current interactions of ultrahigh energy neutrinos with nucleons. This new result is over twice the size of previous estimates and has important implications for cosmic ray experiments now underway as well as for KM3 arrays (cubic kilometer-scale neutrino telescopes) now in the planning stages.

*Introduction* The neutrino-nucleon charged current interaction cross section has a fascinating energy dependence revealing different physical regimes. At the lowest energies, the total cross section goes like the neutrino laboratory energy squared due to the non-relativistic final state phase space. In the next regime, the energy dependence becomes linear, as one expects from Lorentz covariance, dimensional analysis, and the pointlike 4-Fermi interaction. These are sufficient to predict that the cross section goes like ${G_F}^2 s$, where $s$ is the invariant center of mass energy squared. This increase with energy of the cross section has repeatedly been verified and makes the neutrino beam a practical experimental tool. However, the linear rise with energy is unsustainable as it would eventually violate s-wave unitarity. The 4-Fermi model breaks down when $s$ (or the magnitude of the invariant momentum transfer $Q^2$) approaches the W-boson mass squared ${M_W}^2$. Above that energy, the exchange of a single W-boson on an elementary target predicts a cross section which behaves like $\sigma_0 \log(\frac{s}{{M_W}^2})$, where $\sigma_0 = \frac{{G_F}^2 {M_W}^2}{2\pi}$.

This expectation is again revised because the nucleon is a composite, not elementary, object made of quarks and gluons. In the ultra high energy limit, $s \gg {M_W}^2$, the total cross section is dominated by an integration region of nearly constant energy loss, with momentum transfer approximately in the range $0.1{M_W}^2 \leq Q^2 \leq {M_W}^2$, and $x \ll 1$. In this limit the number of quarks with small $x$ thus determines the total cross section. Recent small $x$ data from the H1 and Zeus collaborations at HERA indicate a rather singular growth in the number of quarks carrying small momentum fraction, resulting in a substantial enhancement for the physical nucleon cross section compared with the one found for scattering from an elementary object.

In this Letter we will combine new results from the HERA electroproduction experiments with the total cross section analysis to predict the ultra-high energy (UHE) neutrino cross section at energies of 100 TeV and above. This cross section has important implications for cosmic ray experiments now underway and for KM3 arrays in the planning stages, the so-called neutrino telescopes. Since cosmic ray photons with energies near $10^{15}$ eV are strongly attenuated on the intergalactic photon background, at these energies we are left with neutrinos as the only long lived, light, and electrically neutral elementary particle for viewing the UHE universe. Even at energies where photons are still available, neutrino astronomy opens a new window on the dense, hidden regions of the cosmos which are opaque to photons but virtually transparent to neutrinos [1]. Galactic cores provide good examples of such hidden regions. Calculations of AGN sources [2] predict a sizable flux of UHE neutrinos in the energy range of $10^{14}$ eV and above. Our predictions for the cross section in this range are considerably larger than those of our own previous work (denoted MR) [3], and the subsequent work of Reno and Quigg (RQ) [4], which were both based on incomplete knowledge of small $x$ physics. The situation has changed recently due to HERA electroproduction experiments, where an 800 GeV proton beam and a 30 GeV electron beam is used to study the process $\mathrm{e} + \mathrm{p} \rightarrow \mathrm{e}' + \mathrm{X}$ and extract the quark distribution functions at small $x$. A remarkable rise of the structure functions with decreasing $x$ has been observed. This rise strongly influences the UHE neutrino-nucleon cross section. A microscopic strong interaction process of fundamentally non-perturbative origin (the quark distributions) thus profoundly affects electroweak cosmic rays processes which originate in the most distant parts of the universe.

*Small $x$ Evolution and the Neutrino Cross Section* The kinematics of one-W exchange are standard. The cross section can be written in terms of the scaling variables $x = \frac{Q^2}{2M\nu}$ and $y = \frac{\nu}{E_\nu}$, and the structure functions $\mathcal{F}_1^{\nu N}$ , $\mathcal{F}_2^{\nu N}$ , and $\mathcal{F}_3^{\nu N}$, which depend on $x$ and $Q^2 = sxy$. We use the relation $2x\mathcal{F}_1 = \mathcal{F}_2$ and also ignore terms proportional to $\mathcal{F}_3$, the difference between the number of quarks and antiquarks, which are negligible in the sea dominated region of $x \ll 1$. Thus,

$$\sigma^{\nu N}(s) = \frac{{G_F}^2 {M_W}^2}{2\pi} \frac{{M_W}^2}{s} \int_{\frac{{Q_{\rm low}}^2}{s}}^1 dx \int_{\frac{{Q_{\rm low}}^2}{sx}}^1 dy \frac{(1 - y + \frac{1}{2}y^2)\mathcal{F}_2^{\nu N}(x, sxy)}{(\frac{{M_W}^2}{s} + xy)^2} \,. \tag{1}$$

The denominator of Eq. (1) contains the W propagator $\frac{1}{(Q^2 + {M_W}^2)^2}$ written in terms of $x$ and $y$. The integration is strongly affected by the $\frac{1}{Q^4}$ singularity which is cut off by ${M_W}^2$. The lower integration limits give a slight underestimate of the true cross section and simply reflect ignorance of the structure function in non-perturbative regions. Our procedure is to choose some reasonable $Q^2_{\rm low}$ as a cutoff and then test that the results are insensitive to its precise value. We have found no significant variation over the range $6.4 < Q^2_{\rm low} < 64$ GeV$^2$.

To continue we need the parton distributions in order to form $\mathcal{F}_2^{\nu N} = x[u(x) + \overline{u}(x) + d(x) + \overline{d}(x) + s(x) + \overline{s}(x) + ...]$. Information on these quark distributions can be obtained from electroproduction, where $F_2^{ep} = x[\frac{4}{9}(u(x)+\overline{u}(x)) + \frac{1}{9}(d(x)+\overline{d}(x)) + \frac{1}{9}(s(x)+\overline{s}(x)) + ...]$. The H1 [5] and ZEUS [6] [7] groups at HERA have recently measured $F_2^{ep}$ over a region of $10^{-3.7} < x < 10^{-.96}$ and $10^{0.9} < Q^2 < 10^{3.7}(\mathrm{GeV}^2)$. The important integration region of Eq. (1) extends up to $Q^2 \approx {M_W}^2$, a region where no direct small $x$ parton data on $F_2^{ep}$ exist. Fortunately, QCD evolution via the well known dGLAP [8] evolution equations, together with the available $F_2^{ep}$ data, enable us to effectively bridge this gap.

Our procedure is to fit solutions to the next-to-leading order QCD-dGLAP evolution equations for the parton distributions to the measured structure function $F_2^{ep}$. In the region of $x < 10^{-1}$, these equations have a remarkably simple two parameter solution when the starting distributions have a power-law form in $x$. We will present the details elsewhere. The result is

$$xq(x,Q^2) = Ax^{-\mu}\frac{P_{\rm qg}(\mu+1,Q^2)}{P_{\rm gg}(\mu+1,Q^2)}\exp[\ln\ln(\frac{Q^2}{\Lambda^2_{\rm QCD}})\ln(\frac{Q^2}{\Lambda^2_{\rm QCD}})P_{\rm gg}(\mu+1,Q^2)]\,, \qquad (2)$$

where $P_{\rm qg}(\mu{+}1,Q^2)$ and $P_{\rm gg}(\mu{+}1,Q^2)$ are $\mu{+}1$ moments of the quark-gluon and gluon-gluon splitting functions at leading plus next-to-leading order, and the overall normalization $A$ and the power, $\mu$, are adjustable parameters. Using Eq. (2), we form the structure function with the Ansatz, $q(x) = \overline{q}(x) = u(x) = d(x) = s(x) = 2c(x) = 2b(x)$ so that $F_2^{ep} = \frac{17}{9}xq(x)$ and $\mathcal{F}_2^{\nu N} = 8xq(x)$. This simple form for the structure function gives a high quality fit to the global set of HERA data. Our best fit yields $A = 0.011 \pm 0.002$ and $\mu = 0.40 \pm 0.03$; this result is displayed together with the HERA data in Fig.1. The $\mathcal{X}^2$ per degree of freedom for this fit is 0.85, which is excellent; our quark distributions are in good agreement with existing global fits [9] in the regions where such fits have been reported. Our results are actually in somewhat better overall agreement with the HERA data than any we have seen to date, including those which fit the initial distribution with many parameters.

With a prediction for $\mathcal{F}_2(x, Q^2)$ in hand, we employ Eq. (1) to obtain our cross section. In the "large $x$" region, where the present work has nothing new to add, we use conventional multiple parameter fits for the quark distributions, splicing our result with well established structure function results [9] for $x \geq 0.2$. We find that the cross section for 50 TeV $< E_\nu <$ 100 TeV is not very sensitive to exactly where the splice is made, and it has virtually no effect for $E_\nu >$ 100 TeV. For 50TeV $< E_\nu <$ 50PeV, a compact expression which fits our result is

$$\sigma^{\mu N}(E_\nu) = \exp[-80.56 + 0.822(\ln E_\nu) - 0.0231(\ln E_\nu)^2]\,, \tag{3}$$

with $E_\nu$ in units of TeV. In Fig.2, we compare this new result with our earlier calculation, MR [3], which chose to neglect valence quarks, and also with the calculation of RQ [4], which restored this contribution. Valence effects are small for energies above $\sim$100 TeV. The region of existing high energy neutrino-nucleon cross section data is indicated, along with one very high energy data point recently extracted by the H1 group who measured $\sigma(ep \to \nu + X)$ [10]. The lone data point is at approximately 10 times the highest energy previously reported and is in good agreement with our new result, thus tying together two completely independent aspects of HERA measurements. Varying the position of our $x$ splice over the range $0.4 > x > 0.01$ changes the low energy portion ($<$ 100 TeV) of our result by less than the error bars on the H1 data point. Our cross section is roughly a factor of 2.2 larger than previous estimates over the range 100 TeV $< E_\nu <$ 10 PeV.

At higher energies the rate of growth of our cross section is larger because the small $x$ parton distributions are following a "power law" evolution that self-generates an accelerated $Q^2$ evolution. It is important to note that this is an *a-priori* unknown, non-perturbative boundary condition for the evolution, entirely consistent with the perturbative QCD formalism used to study it. Differences between the new and old physics are entirely in this boundary condition. Previous formulas were based on $\exp(\sqrt{\ln\frac{1}{x}\ln\ln Q^2})$ evolution of a parton beginning at $x = 1$ and creating a shower (the Green function of dGLAP evolution). We note that the question of parton saturation or recombination, which has received much attention at the lower $Q^2$ regions accessible to the HERA experiments, is a "higher twist" effect (subdominant to all powers of logarithms) and not relevant at the rather large $Q^2 \approx O({M_W}^2)$ we study. Using our distributions self consistently, the highest energy (smallest $x$) that can safely be considered are $E_\nu < 10^3$ PeV ($x \geq 10^{-7}$).

*Implications for UHE Neutrino Astronomy* High-energy neutrino telescopes are multipurpose detectors that are expected to make important contributions in astrophysics, and possibly particle physics. Conventional sources of UHE neutrinos include scattering of $\sim 10^{20}$ eV protons off of 3°K background photons; these events produce charged pions which then decay to produce neutrinos. The Fly's-Eye collaboration has reported detailed evidence for correlated changes in the spectrum and composition of cosmic rays in the region above $10^{18}$ eV [11]. Expected point sources within our own galaxy include X-ray binaries such as Hercules X-1, Cygnus X-3, and the Crab Nebula. Recently, much attention has been given to active galactic nuclei (AGN) as sources for the highest energy cosmic rays and large fluxes of UHE neutrinos [12]. Learned and Pakvasa [13] have recently discussed the "new physics" signal of tau neutrino oscillations at PeV energies.

A natural "KM3" scale for UHE neutrino telescopes is 1 km$^3$. Several current prototypes covering a few percent of this volume are now in various stages of deployment. The DUMAND, NESTOR and BAIKAL experiments are located in deep ocean water, deep lake water and shallow lake water respectively while the antarctic AMANDA project is situated under ice. These experiments detect Cherenkov radiation at optical frequencies. There is also a realistic possibility of using coherent radio techniques [14] to enhance detection for energies >50-100 TeV.

As an example of what our UHE cross section implies for KM3 telescopes, we present an event rate calculation for 1 km$^3$ of polar ice based on the latest estimates of UHE neutrino flux from AGN sources by Szabo and Protheroe [2]. In general, a larger cross section translates into a higher event rate. However, if one is interested in upward going (through the earth) neutrinos, then the competing effect of neutrino absorption by the earth must be carefully considered. An equivalent solid angle can be calculated which expresses the fraction of the original flux which actually reaches the detector [15]. For our calculation we made use of the PREM earth structure model [16] from which an integrated nucleon density over possible incident neutrino directions can be obtained. Our result appears as part (a) of Fig. 3, where the effect of our larger cross section can be seen by comparison with results obtained using the RQ cross section. The energy at which 50% of the upward flux fails to reach the detector is about 700 TeV in the case of RQ compared with roughly 180 TeV using our results.

In part (b) of Fig. 3 we show a calculation of the differential event rate, the number of electromagnetic cascades per year per TeV per cubic kilometer of ice, induced by muon neutrinos. Rates for upward and downward going events are indicated separately and dashed versus solid lines indicate results using the two cross sections as in part (a). Due to the effect of earth shadowing, the upward rates are always smaller than downward rates with the difference becoming more pronounced as cascade energy increases. Interestingly, the upward rates for our larger cross section are higher than for the MR or RQ cross sections despite the lower flux at the detector due to shadowing. In the competition between attenuated flux versus increased interaction probability for the surviving particles, the net effect is an increase in rate.

Rate predictions for a specific experimental situation depend on an effective detection volume for the apparatus as a function of the event (cascade) energy, which include array geometry, receiver characteristics, signal attenuation in the medium, noise characteristics and detection threshholds. Given the bulk event rates shown in Fig. 3, the prospects for a KM3-scale array of almost any reasonable design are very encouraging as is the corresponding potential for scientific discovery.

## ACKNOWLEDGMENTS

This work was supported in part under Department of Energy Grant Number DE-FGO2-85-ER 40214 and by the *Kansas Institute for Theoretical and Computational Science* through the K*STAR/NSF program. DWM thanks the Department of Physics at the University of California, Davis, and in particular Barry Klein, Ling-Lie Chau and Jack Gunion, for hospitality during the course of this work. Finally, we wish to thank Malcolm Derrick (ANL) for keeping us up to date with the latest measurements from ZEUS.

## FIGURES

FIG. 1. Our best fit (lines) compared with the H1 and ZEUS data. Numbers just above each line indicate $Q^2$ values for the data in $\mathrm{GeV}^2$. For clarity, the data and associated fits are separated by integer powers of 3 as $Q^2$ increases. Thus, we show $3^0$ times $\mathcal{F}_2(x, 8.5)$, $3^1$ times $\mathcal{F}_2(x, 12-15)$, and so on thru $3^{10}$ times $\mathcal{F}_2(x, 5000)$.

FIG. 2. Total charged current neutrino-nucleon cross section versus incident neutrino energy. The extrapolated 4-Fermi result is shown along with the single high energy HERA data point. Also shown is the present ultrahigh energy prediction (solid line) along with previous results of MR and RQ.

FIG. 3. (a) Equivalent solid angle (units of $2\pi$) for the upward going neutrino flux showing the increasing opacity of the earth with energy. Attenuation is greater with the new larger cross section (denoted FMR) compared with the RQ cross section estimate. (b) The differential event rate per year per TeV versus muon energy for a cubic kilometer of polar ice due to AGN neutrinos. We compare results obtained using the FMR and RQ cross sections with separate curves for upward versus downward going events. Note that the integrated rate between $10^2$ TeV and $10^3$ TeV for upward going events using the new cross section is 8029 events/year/$\mathrm{km}^3$.